\documentclass[english,10pt,final,twocolumn]{elsarticle}
\usepackage[T1]{fontenc}
\usepackage[latin9]{inputenc}
\usepackage{pdfpages}
\usepackage{color,soul}
\usepackage{amstext}
\usepackage{graphicx}
\usepackage{overpic}
\makeatletter

\newcommand{\lyxmathsym}[1]{\ifmmode\begingroup\def\b@ld{bold}
  \text{\ifx\math@version\b@ld\bfseries\fi#1}\endgroup\else#1\fi}
\usepackage{braket}
\usepackage{mathrsfs}
\usepackage{slashed}
\usepackage{bm}
\usepackage{datetime}
\usepackage{siunitx}
\usepackage{placeins}
\DeclareSIUnit[number-unit-product = {}]\clight{c}
\DeclareSIUnit\eVperc{\eV\per\clight}
\DeclareSIUnit\GeVpercs{\giga\eV\squared\per\clight\squared}
\DeclareSIUnit\MeVpercs{\mega\eV\per\clight\squared}
\journal{Physics Letters B}
\biboptions{numbers,sort&compress}
\usepackage[left=1.6cm,right=1.04cm,top=1.65cm,bottom=1.65cm,columnsep=25pt]{geometry}
\usepackage{lineno}
\usepackage{hyperref}  % hyper-links 
\hypersetup{breaklinks = true, colorlinks = true, allcolors = magenta}
\usepackage{setspace}
\usepackage{graphicx}
\usepackage{wrapfig}

\makeatother

\usepackage{babel}
\begin{document}

\begin{frontmatter}{}

\title{The influence of Fermi motion on the comparison of the polarization
transfer to a proton \\in elastic $\vec ep$ and quasi-elastic $\vec eA$ scattering}

\author[TAU]{S.~Paul\corref{cor2}}

%\ead{davidizraeli@post.tau.ac.il}
\ead{paulsebouh@mail.tau.ac.il}

\author[JSI]{T.~Brecelj}

\author[Mainz]{H.~Arenh\"ovel}

\author[Mainz]{P.~Achenbach}

\author[TAU]{A.~Ashkenazi}

\author[JSI]{J.~Beri\v{c}i\v{c}}

\author[Mainz]{R.~B\"ohm}

\author[zagreb]{D.~Bosnar}

\author[TAU]{E.O.~Cohen}

\author[JSI]{L.~Debenjak }

\author[Mainz]{M.O.~Distler}

\author[Mainz]{A.~Esser}

\author[zagreb]{I.~Fri\v{s}\v{c}i\'{c}\fnref{mit}}

\author[Rutgers]{R.~Gilman}

\author[TAU]{D.~Izraeli}

\author[JSI]{T.~Kolar}

\author[TAU,nrc]{I.~Korover}

\author[TAU]{J.~Lichtenstadt}

\author[TAU,soreq]{I.~Mardor}

\author[Mainz]{H.~Merkel}

\author[Mainz]{D.G.~Middleton}

\author[JSI,Mainz,UL]{M.~Mihovilovi\v{c} }

\author[Mainz]{U.~M\"uller}

\author[TAU]{M.~Olivenboim}

\author[TAU]{E.~Piasetzky}

\author[Mainz]{J.~Pochodzalla}

\author[huji]{G.~Ron}

\author[Mainz]{B.S.~Schlimme}

\author[Mainz]{M.~Schoth}

\author[Mainz]{F.~Schulz}

\author[Mainz]{C.~Sfienti}

\author[UL,JSI]{S.~\v{S}irca}

\author[JSI]{S.~\v{S}tajner }

\author[USK]{S.~Strauch}

\author[Mainz]{M.~Thiel}

\author[Mainz]{A.~Tyukin}

\author[Mainz]{A.~Weber}

\author[TAU]{I.~Yaron}

\author{\\\textbf{(A1 Collaboration)}}

\cortext[cor2]{Corresponding author}

\fntext[mit]{Present address: MIT-LNS, Cambridge, MA 02139, USA.}

\address[TAU]{School of Physics and Astronomy, Tel Aviv University, Tel Aviv 69978,
Israel.}

\address[JSI]{Jo\v{z}ef Stefan Institute, 1000 Ljubljana, Slovenia.}

\address[Mainz]{Institut f\"ur Kernphysik, Johannes Gutenberg-Universit\"at, 55099
Mainz, Germany.}

\address[zagreb]{Department of Physics, Faculty of Science, University of Zagreb, HR-10000 Zagreb, Croatia.}

\address[Rutgers]{Rutgers, The State University of New Jersey, Piscataway, NJ 08855,
USA.}

\address[nrc]{Department of Physics, NRCN, P.O. Box 9001, Beer-Sheva 84190, Israel.}

\address[soreq]{Soreq NRC, Yavne 81800, Israel.}

\address[UL]{Faculty of Mathematics and Physics, University of Ljubljana, 1000
Ljubljana, Slovenia.}

\address[huji]{Racah Institute of Physics, Hebrew University of Jerusalem, Jerusalem
91904, Israel.}

\address[USK]{University of South Carolina, Columbia, SC 29208, USA.}

\begin{abstract}
A comparison between polarization-transfer to a bound proton in quasi-free kinematics by the
A$(\vec{e},e'\vec p)$ knockout reaction and that in elastic scattering off a free proton can provide information on the
characteristics of the bound proton.
In the past the reported measurements have been compared to those of a free proton with zero initial momentum. We introduce, for the first time, expressions for the polarization-transfer components when the proton is initially in motion and compare them to the $^2$H data
measured at the Mainz Microtron (MAMI).  We show the ratios of the transverse ($P_x$) and longitudinal ($P_z$) components of the polarization transfer in
$^2\textrm{H}(\vec{e},e'\vec p)\textrm{n}$, to those of elastic scattering off a ``moving proton'', assuming the proton's initial (Fermi-motion) momentum equals the negative missing momentum in the measured reaction.  We found that the correction due to the proton motion is up to 20\% at high missing momentum.  
 However the effect on the double ratio $\frac{(P_x/P_z)^A}{(P_x/P_z)^{^1\!\textrm{H}}}$ is largely canceled out, as shown for both $^2$H and $^{12}$C data.  
 %This implies that the kinematics is not the primary cause for the deviations between quasi-elastic and elastic scattering reported previously.  
 This implies that the difference between the resting- and the moving-proton kinematics is not the primary cause for the deviations between quasi-elastic and elastic scattering reported previously.

\end{abstract}
\date{\today}

\end{frontmatter}{}
%define what subscript to use for the moving proton kinematics
\newcommand{\mov}{M}
\section{Introduction}
Polarization transfer from a polarized electron to a proton in an elastic scattering reaction has
become a recognized method to measure the ratio of the proton's elastic electromagnetic form factors
$G_E/G_M$ \cite{Jones:1999rz,Gayou:2001qd,Punjabi:2005wq,Milbrath:1997de,Barkhuff:1999xc,Pospischil:2001pp,PhysRevC.64.038202,MACLACHLAN2006261,PhysRevC.74.035201}.  For a proton initially at rest, assuming the one-photon exchange approximation, the ratio of the transverse ($P_x$)
to longitudinal ($P_z$) polarization-transfer components, with respect to the momentum transfer $\vec q$, is proportional to $G_E/G_M$ \cite{Akh74}:
\begin{equation}
\left(\frac{P_x}{P_z}\right)_{\rm H} =  -\frac{2M}{(k_0+k_0')\tan (\theta_{k,k'}/2)}\frac{G_E}{G_M},
\label{eq:Rs}
\end{equation}
where $k_0$, $k^{\,\prime}_0$ are respectively the initial and final electron energies, $\theta_{k,k'}$ is the electron scattering angle in the lab frame, and $M$ is the proton's mass. This provides a direct measurement of the form factor (FF) ratio and eliminates many systematic uncertainties \cite{Perdrisat}.

Measuring the ratio of the components of the polarization transfer to a knock-out proton in quasi-free
kinematics on nuclei, which is sensitive to the electromagnetic FF ratio,
has been suggested as a method to study differences between free and bound protons  \cite{Milbrath:1997de,Barkhuff:1999xc}.
As such it can be used as a tool to identify
medium modifications in the bound proton's internal structure \cite{Sargsian}, reflected in the FFs and thereby in the
polarization transfer.  Deviations between polarization ratios in quasi-free and elastic scattering can be interpreted only with realistic calculations of the nuclear effects, such as final state interactions (FSI), meson exchange currents (MEC), isobar currents (IC), and relativistic corrections on the outgoing proton
polarization components \cite{deep2012PLB,deepCompPLB,Arenhovel}.  However, a comparison to the polarization transfer on a free proton should also consider the Fermi motion of the struck proton, rather than comparing to a reaction with the proton at rest. 

Polarization-transfer experiments have been carried out recently on $^2$H and $^{12}$C target nuclei at the
Mainz Microtron (MAMI) \cite{deep2012PLB,deepCompPLB,ceepLet}, as well as on $^2$H, $^4$He and $^{16}$O at Jefferson Lab \cite{jlabDeep,Strauch,Paolone,Dieterich,Malov_O16}, in search of medium modification in the bound proton's internal structure. In
particular, these experiments were performed to study deeply bound nucleons, characterized by high
missing momentum, which is equivalent (neglecting FSI) to protons with high initial momentum.  It was shown in \cite{deep2012PLB,deepCompPLB} that for $^2$H at low momentum transfer\footnote{These data were taken at $Q^2$ = 0.18 and 0.4 (GeV/$c$)$^2$.}, the deviations can be explained by
nuclear effects without the necessity of introducing modified FFs.  In these experiments, the comparison to the free proton  was done either by measurements of the polarization
transfer ratio to $^1$H at similar kinematics \cite{Strauch,jlabDeep}, or by calculations using Eq.\ \ref{eq:Rs} and a fit to the world data of proton FFs \cite{Perdrisat,Bernauer}, which were used in \cite{deep2012PLB,deepCompPLB,ceepLet}.  

Kinematically, quasi-elastic scattering differs from elastic scattering in that the bound proton is \textit{off}-shell, and in that it moves relative to the nucleus with Fermi motion.  
However, using Eq. \ref{eq:Rs} to calculate the polarization ratio is only valid if the proton is both \textit{on}-shell and at rest.  

In this work we introduce an alternative approach: 
comparing the polarization transfer in quasi-free scattering
to that of a free proton (on-shell) with a finite initial momentum.  
The general expressions for polarization transfer to a free moving proton have been developed.
We applied this prescription to the $^2$H and $^{12}$C data measured at MAMI over relatively large missing momentum ranges. 
 The elastic kinematics are described in Section \ref{sec:kinematics}, while the polarization transfer formulae are given in Section \ref{sec:pol_trans}.  The application of this prescription to MAMI data is shown in Section \ref{sec:application}.

\section{Kinematics}
\label{sec:kinematics}
The kinematics are defined by an electron with initial four-momentum $k$ scattering off a proton with initial four-momentum $p$, through exchange of a virtual photon with four-momentum $q$, resulting in the two particles having final four-momenta $k'$ and $p'$ respectively.  

The final momenta of the proton and electron, $\vec p^{\,\prime}$ and $\vec k^{\,\prime}$, are measured in the spectrometers.   The ``missing momentum'', defined as $\vec{p}_{\rm miss}=\vec{q}-\vec p^{\,\prime}$, is calculated using the momentum transfer defined by $\vec{q}=\vec{k}-\vec k ^{\,\prime}$.   In the absence of FSI, the initial proton momentum is given by the missing momentum, $\vec{p}=-\vec{p}_{\rm miss}$.  By convention, the missing momentum is considered positive (negative) if $\vec p_{\rm miss}\cdot\vec q$ is positive (negative).  
%We refer to the missing momentum as being positive (negative) if the component of $\vec{p}_{\rm miss}$ in the direction of $\vec{q}$ is positive (negative).  

\begin{figure}[th]
\begin{center}
\includegraphics[width=.63\columnwidth]{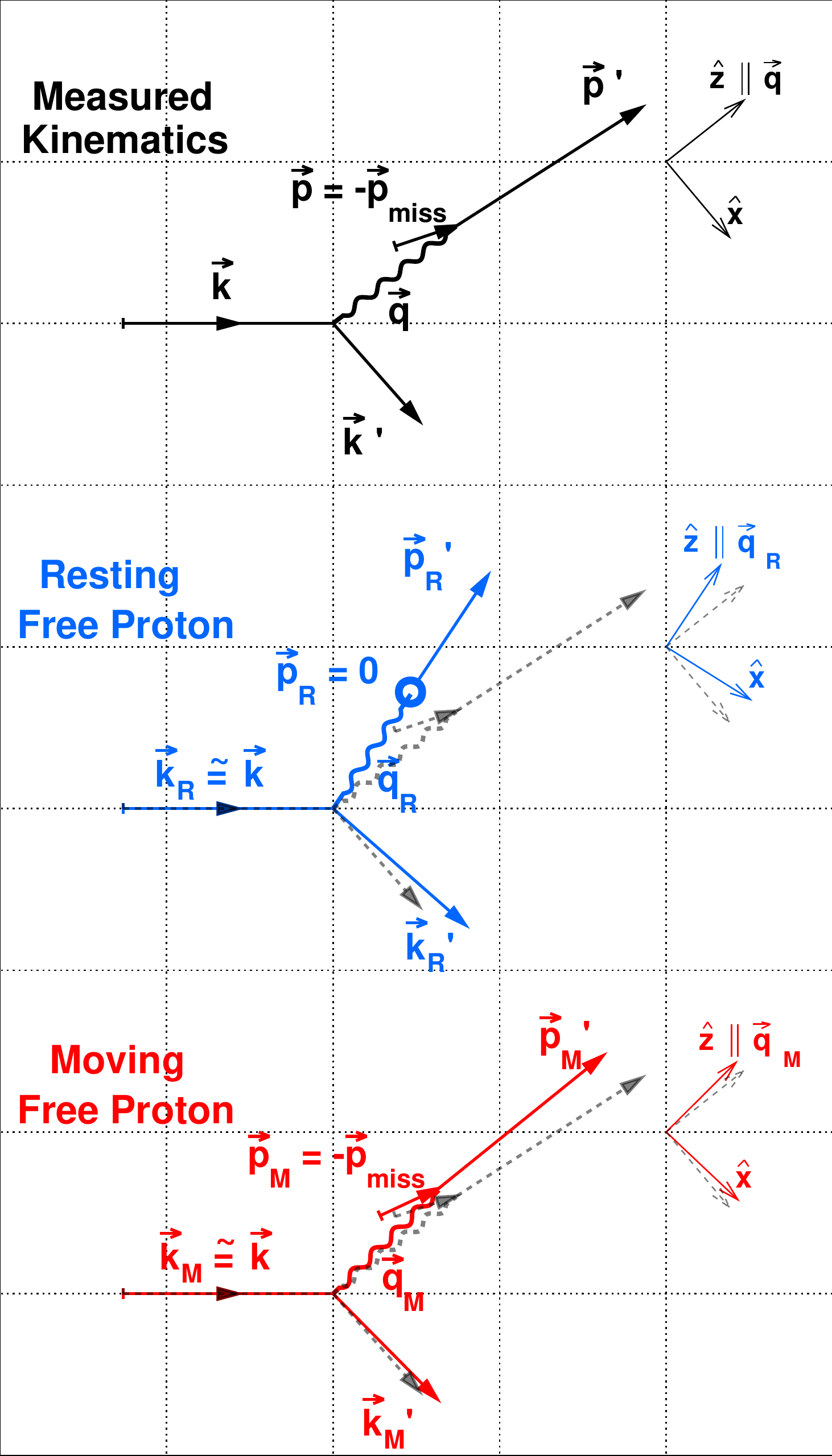}
\caption{
The measured kinematics for a sample QE event (top) compared to the resting (middle) and moving (bottom) elastic kinematics.  In the middle and bottom plots, the solid lines show the free-proton kinematics and the dashed lines show the measured kinematics.  In the middle plot, the initial momentum of the proton is zero.  In the bottom plot, the proton's initial momentum is the negative missing momentum, which neglecting FSI, is equal to the bound proton's initial momentum $\vec{p}$ in the top plot.  The transformation from the measured kinematics to the moving free-proton kinematics causes smaller changes to the magnitudes of the vectors $\vec{k}$, $\vec k^{\,\prime}$, $\vec{q}$ and $\vec p^{\,\prime}$, and the angles between them, than the transformation to resting free-proton kinematics.
}
\label{fig:kinematics_comparison}
\end{center}
\end{figure}

These measured kinematics are shown in the top diagram in Figure \ref{fig:kinematics_comparison}, and are compared, in the middle and bottom diagrams, to the kinematics of elastic scattering off a proton initially at rest, and off a proton with initial momentum $\vec p = -\vec p_{\rm miss}$, respectively.  %with a resting and a moving proton, respectively.

Each measured quasi-elastic event is compared to an elastic event that
has the same incident energy ($k_0$), the same magnitude of the four-momentum transfer ($Q^2 = -q^2$) and the same initial proton momentum ($\vec{p}$) as the measured quasi-elastic event.  Also, the struck proton is on-shell in its initial and final state in the elastic event\footnote{This is equivalent to the Bjorken condition, $\left[\frac{Q^2}{2p\cdot q}\right]_\mov=1$.}.  These criteria uniquely define a set of kinematics, here referred to as ``moving-proton'' kinematics. We use a subscript ``$_\mov$'' to distinguish between the kinematic variables in the moving-proton kinematics from the measured quantities.   %The kinematic variables in these kinematics are distinguished from the measured values by a subscript $\mov$.

\FloatBarrier

In the standard coordinate system for polarization transfer, where $\vec{q}\,||\,\hat{z}$ and the scattering is in the $xz$ plane, the moving-proton kinematics are\footnote{For the derivation, see the supplementary material.}:

\begin{equation}
\label{eq:kinematics_mov_start}
p_\mov = \left(p_{0\mov}, \;p_x, \;p_y, \;p_z \right),
\end{equation}
\begin{equation}
k_\mov = \left(k_0, \; \frac{Q\sqrt{4k_0(k_0-\omega_\mov)-Q^2}}{2q_{z\mov}}, \; 0 ,\; \frac{2k_0\omega_\mov+Q^2}{2q_{z\mov}}\right),
\end{equation}
\begin{equation}
p'_\mov = p_\mov+q_\mov,
\end{equation}
and 
\begin{equation}
k'_\mov = k_\mov - q_\mov,
\end{equation}
where
\begin{equation}
q_\mov = \left(\omega_\mov, \;0, \;0, \;q_{z\mov} \right),
\end{equation}
\begin{equation}
 \omega_\mov =  \frac{Q^2 p_{0\mov}+p_z Q\sqrt{4E_{pt}^2+Q^2}}{2E_{pt}^2},
 \label{eq:omega_M}
 \end{equation}
 and 
 \begin{equation}
 q_{z\mov} = \sqrt{Q^2+\omega_\mov^2},
\label{eq:kinematics_mov_end}
 \end{equation}
 using $p_{0\mov}$ and $E_{pt}$ as shorthand for $\sqrt{M^2+|\vec p |^2}$ and $\sqrt{M^2+p_x^2+p_y^2}$, respectively.  Note that $p_x$, $p_y$ and $p_z$ are the coordinates of $\vec{p}=-\vec{p}_{\rm miss}$ in the $\vec{q}\,||\,\hat{z}$ coordinate system.

In previous publications \cite{deep2012PLB,deepCompPLB,ceepLet}, we used a different type of free-proton kinematics (``resting proton'', denoted with a subscript ``$_R$'' in Fig.~\ref{fig:kinematics_comparison} and elsewhere in this paper) where instead of having the same $\vec{p}$ as in the measured kinematics, we had $\vec{p}_R =\vec{0}$.  The resting-proton kinematics may therefore be evaluated by substituting $\vec{p} = 0$ in Eqs.~\ref{eq:kinematics_mov_start}-\ref{eq:kinematics_mov_end}.

%\FloatBarrier
\section{Polarization transfer}
\label{sec:pol_trans}
The general expressions for the polarization transfer from a longitudinally polarized electron to an initially moving nucleon with momentum $\vec{p}$ in the scattering plane are presented below, with more details in \cite{ArenhovelMoving}: 

\begin{equation}
P_x = -C_PG_M\left(G_E (\alpha_{\ell x} - \beta_{\ell x}) + G_M \beta_{\ell x}\right),
\label{eq:Px}
\end{equation} 
\begin{equation}
P_z = -C_PG_M\left(G_E (\alpha_{\ell z} - \beta_{\ell z}) + G_M \beta_{\ell z}\right),
\label{eq:Pz}
\end{equation} 
where 
\begin{equation}
C_P = \frac{2}{(1+\tau)M^2\Sigma_0},
\end{equation} 
\begin{equation}
\Sigma_0 = (G_E^2+\tau G_M^2)\left(\frac{(K\cdot p)^2}{M^2Q^2(1+\tau)}-1\right)+2\tau G_M^2,
\end{equation} 
\begin{equation}
\label{eq:alxetc}
\alpha_{\ell x} = \frac{1+\tau}{2} \left(\frac{p_x}{p'_0+M}\left(MK_0+2p\cdot k-\frac{Q^2}{2}\right) - MK_x \right),
\end{equation}
\begin{equation}
\beta_{\ell x} = \frac{-1}{4M^2}\left(2k\cdot p - \frac{Q^2}{2}\right)\frac{p_x}{p'_0+M}\left(M\omega-\frac{Q^2}{2}\right),
\end{equation}
\begin{equation}
\alpha_{\ell z} = \frac{1+\tau}{2} \left(\frac{p_z+q_z}{p_0'+M}\left(MK_0+2p\cdot k-\frac{Q^2}{2}\right)
- MK_z \right),
\end{equation}
and
\begin{equation}
\beta_{\ell z} = \frac{-1}{4M^2}\left(2k\cdot p - \frac{Q^2}{2}\right)\left(\frac{p_z+q_z}{p_0'+M}\left(M\omega-\frac{Q^2}{2}\right)-Mq_z\right),
\end{equation} 
using $\tau = \frac{Q^2}{4M^2}$ and $K = k+k'$.
The factors of $-C_PG_M$ in Eq.\ \ref{eq:Px} and \ref{eq:Pz} cancel out in the ratio:
\begin{equation}
\label{eq:PxPz_HA}
\frac{P_x}{P_z} = \frac{G_E (\alpha_{\ell x} - \beta_{\ell x}) + G_M \beta_{\ell x}}{G_E (\alpha_{\ell z} - \beta_{\ell z}) + G_M \beta_{\ell z}}.
\end{equation}

Equation \ref{eq:PxPz_HA} reduces to Equation \ref{eq:Rs} if the proton is initially at rest. It should be noted that the formulae in this section are only valid for \textit{elastic} scattering in the one-photon approximation.  

\section{Application to MAMI A$(\vec{e},e'\vec p)$ data}
\label{sec:application}
Polarization-transfer measurements on $^2$H and $^{12}$C have been carried out at MAMI \cite{a1aparatus,Pospischil:2000pu} at
momentum transfers of $Q^2$ = 0.18 and 0.40 (GeV/c)$^2$ \cite{deep2012PLB,deepCompPLB,ceepLet}. The kinematic settings of the
measurements are given in Tables 1 and 2 of the supplemental material. The double ratios $\frac{(P_x/P_z)^A}{(P_x/P_z)^{^1\!\textrm{H}}}$ were presented for both nuclei  as functions of the struck proton's virtuality (``off-shellness''), $\nu\equiv(p'-q)^2-M^2$, which
proved to be a useful parameter for a unified description of the deviation of the
bound proton from a free one. 
In addition, Ref.~\cite{deepCompPLB} presents the ratios $\frac{(P_x)^A}{(P_x)^{^1\!\rm H}}$ and $\frac{(P_z)^A}{(P_z)^{^1\!\rm H}}$ for the deuteron data, and showed that they were in agreement with the theoretical calculation, indicating that the deviations from the free proton were primarily due to FSI.

\begin{figure}[h!]
\begin{center}
\includegraphics[width=.97\columnwidth]{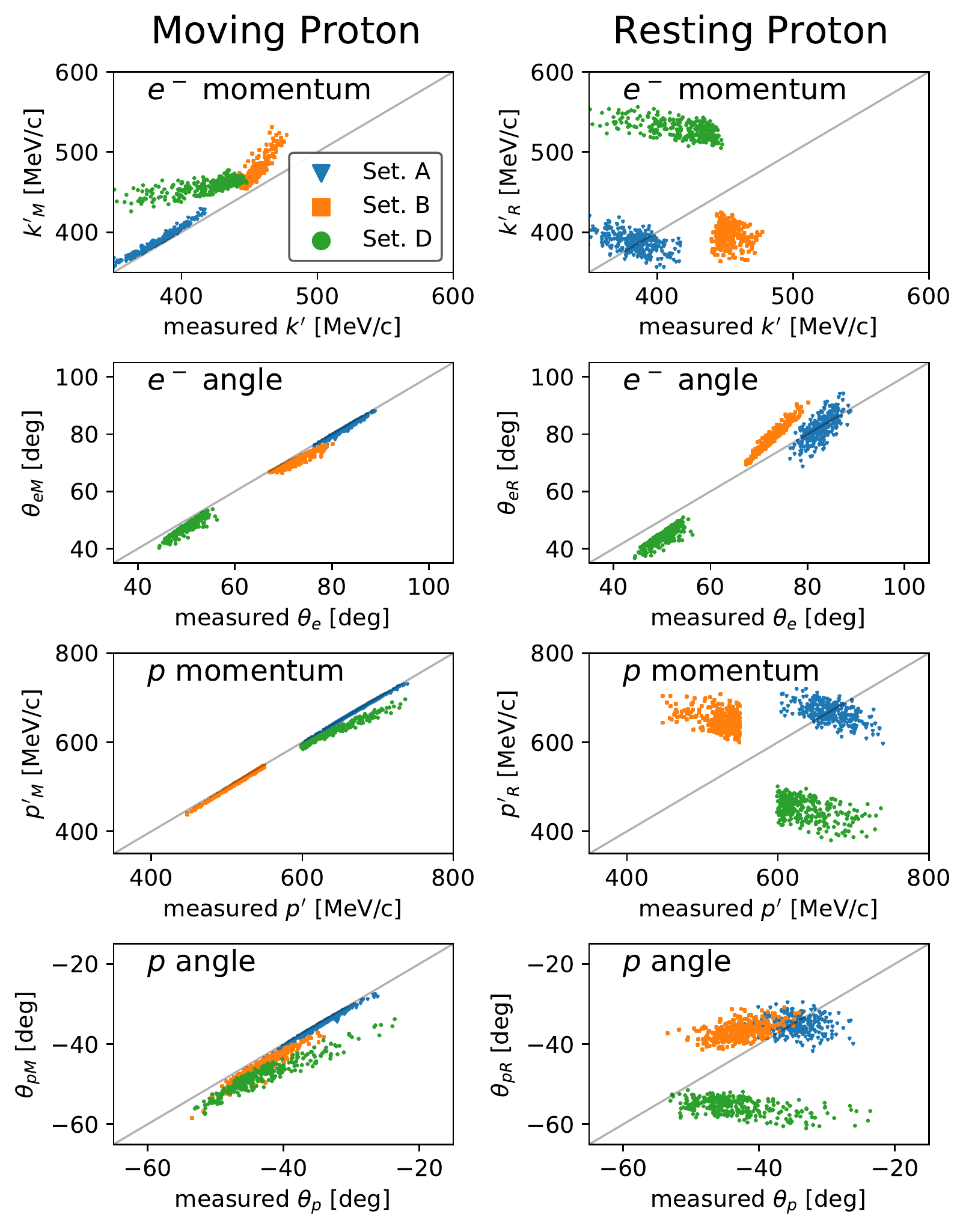}
\caption{
Comparison of the kinematic variables measured for $^2$H in \cite{deep2012PLB,deepCompPLB} ($x $-axis) to those in the moving-proton (left) and resting-proton (right) prescriptions ($y$-axis).  The different colors represent different kinematic settings.  The variables, from top to bottom, are the momentum and angle of the scattered electron ($k'$ and $\theta_e$),  and those of the recoiling proton ($p'$ and $\theta_p$).  The two angles are measured relative to the beam $\vec k$.  These variables were selected because they are the ones that are directly measured in the experiment.  
}
\label{fig:variables}
\end{center}
\end{figure}

Applying the moving-proton prescription to the deuteron data \cite{deep2012PLB,deepCompPLB}, we notice a good correspondence between the measured and evaluated kinematic variables in the left side of Fig.~\ref{fig:variables}.  These differences are much smaller than those between the \textit{resting}-proton kinematics and the measured kinematics, shown on the right side of Fig.~\ref{fig:variables}. 

\FloatBarrier
\FloatBarrier

The differences between the measured and the resting-proton kinematics are more pronounced for settings B and D than in setting A, due to larger $|\vec p_{\rm miss}|$ and virtuality in the former.  

There are larger differences between the moving-proton kinematics and the measured kinematics in the $^{12}$C data \cite{ceepLet}, as shown in the supplementary material, due to the larger Fermi motion in the carbon nucleus.  However, they are still much smaller than the difference between the resting-proton kinematics and the measured kinematics.  

\begin{figure}[th]
\includegraphics[width=.97\columnwidth]{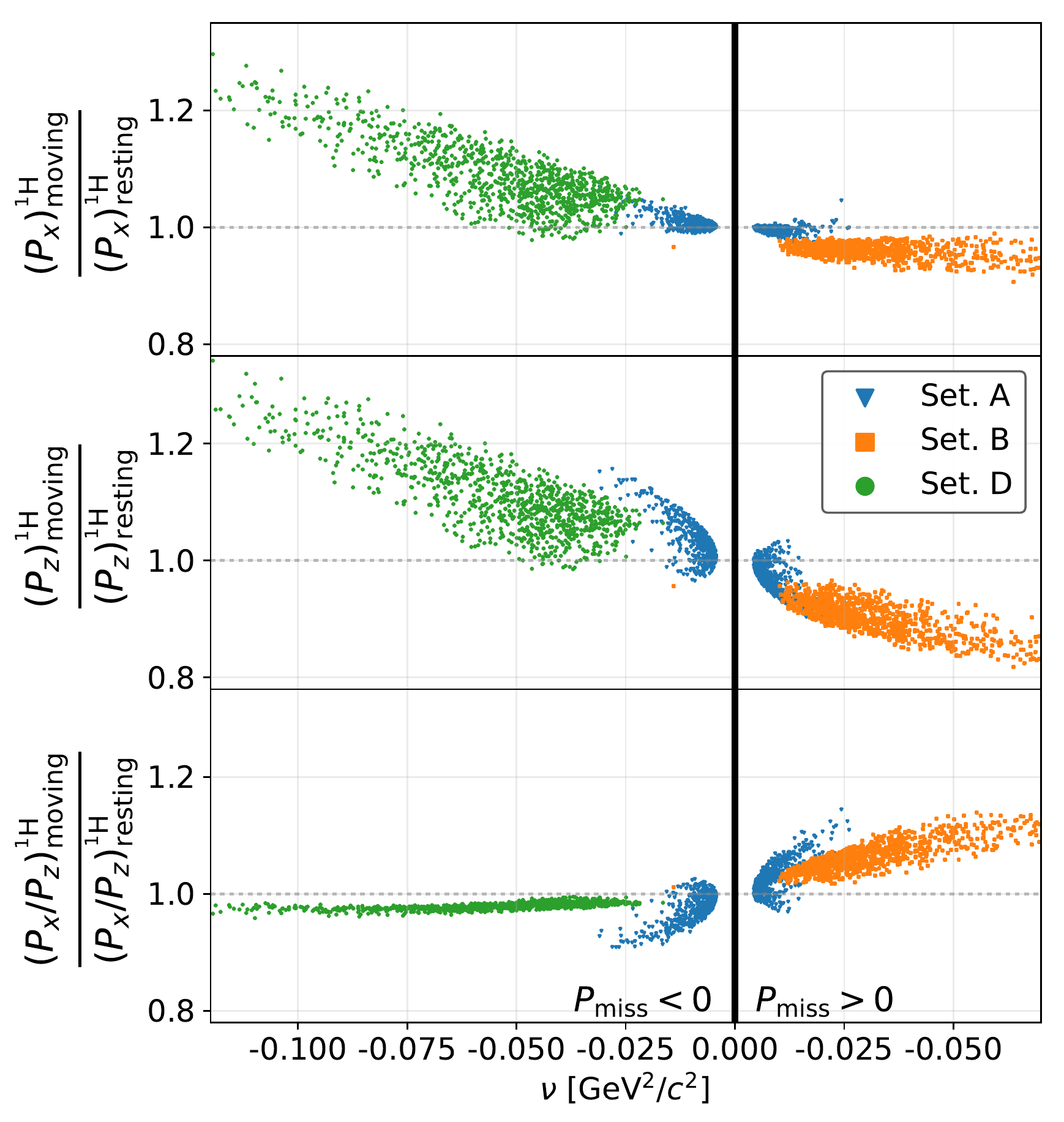}
\caption{Ratios of $P_x$, $P_z$ and $P_x/P_z$ for moving free protons to those for free resting protons, using the kinematics derived from measured events in $^2\textrm{H}(\vec{e},e'\vec p)\textrm{n}$ reactions reported in \cite{deepCompPLB}.  Different colors represent different kinematic setups.  See \cite{deep2012PLB,deepCompPLB} for details.}
\label{fig:corrections}
\end{figure}

In Figure \ref{fig:corrections}, we compare the polarization-transfer components calculated for the resting- and the
moving-proton kinematics.  We show in this figure the ratios $(P_x)^{^1\!\rm H}_{\rm moving}/(P_x)^{^1\!\rm H}_{\rm resting}$, $(P_z)^{^1\!\rm H}_{\rm moving}/(P_z)^{^1\!\rm H}_{\rm resting}$, and $(P_x/P_z)^{^1\!\rm H}_{\rm moving}/(P_x/P_z)^{^1\!\rm H}_{\rm resting}$,
as calculated for the
kinematics of the $^2$H events in Ref.~\cite{deep2012PLB}. The difference between resting and moving for
the individual components is small around $\nu = 0$,  and it increases up to 20\% (15\%) at large virtuality\footnote{This is expected, since the $|\vec p_{\rm miss}|$ is near zero at small $\nu$ and increases monotonically with $\nu$ for any given target and residual nuclear masses.} 
with negative (positive) $p_{\rm miss}$.
The difference between moving and resting in the
ratio $P_x/P_z$ is small: up to 3\% for $p_{\rm miss} < 0$ and up to 10\% at $p_{\rm miss} > 0$.
 
\begin{figure}[th]
\includegraphics[bb=0bp 9bp 471bp 520bp,clip,width=1\columnwidth]{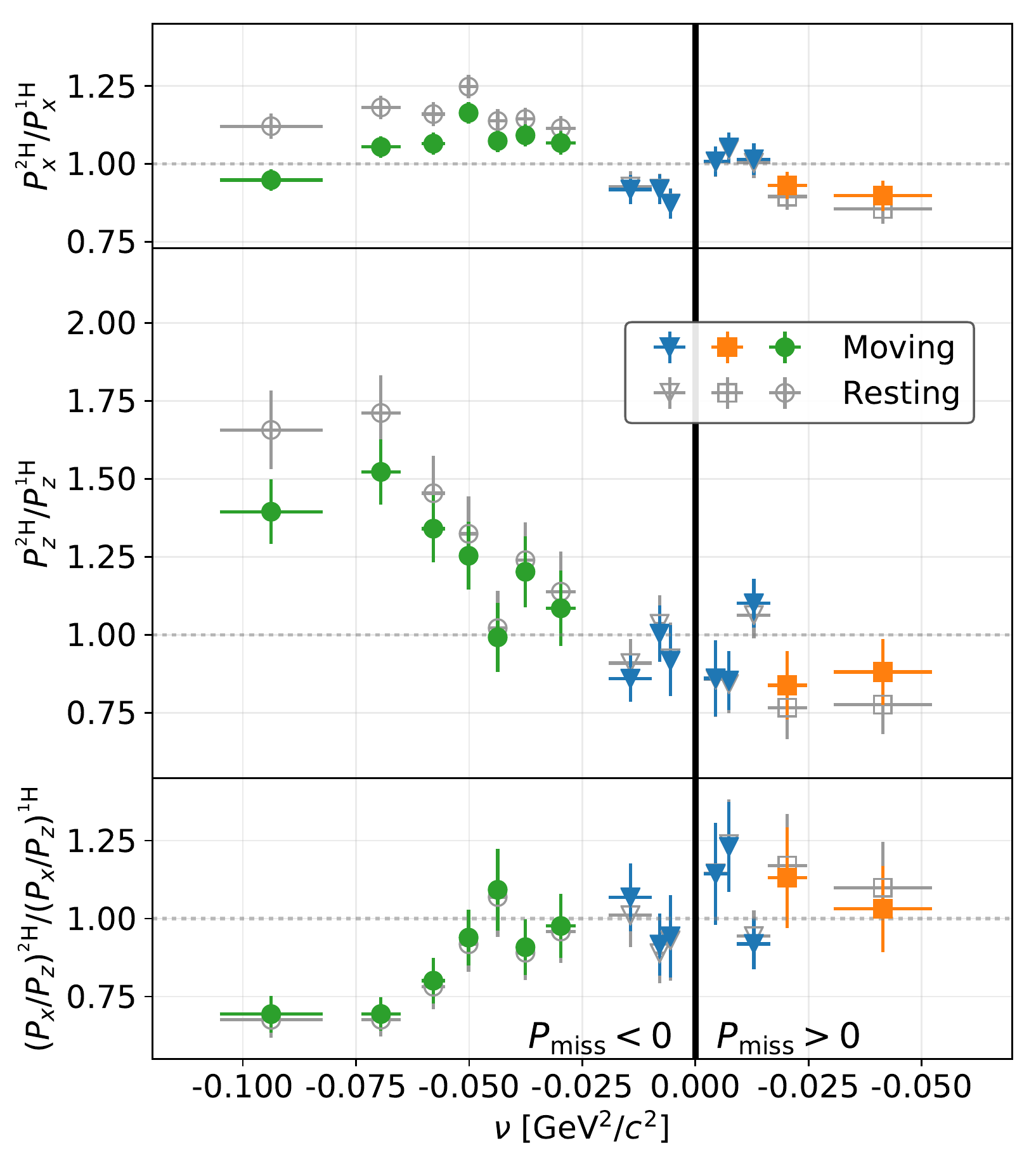}
\caption{The measured ratios, $(P_x)^{^2\rm H}/(P_x)^{^1\rm H}$ and $(P_z)^{^2\rm H}/(P_z)^{^1\rm H}$ and the double ratio
$(P_x/P_z)^{^2\rm H}/(P_x/P_z)^{^1\rm H}$, using \textit{moving} free-proton in the denominator are shown as functions of the proton virtuality, $\nu$ (filled colored symbols).  
These are compared to the same ratios with \textit{resting} free-proton kinematics (grey open symbols) from \cite{deepCompPLB}.  The virtuality dependence is shown separately for positive and negative missing momenta. The different symbols and colors for the data of this work correspond to the different kinematical settings. }
\label{fig:deuteronComponents}
\end{figure}

The ratios of the polarization observables measured for the deuteron to the values calculated event-by-event for a moving proton using Equations \ref{eq:Px} and \ref{eq:Pz} are shown in Figure \ref{fig:deuteronComponents}.  These are contrasted with the corresponding ratios for the resting proton, which were reported in \cite{deepCompPLB}.  
The double ratio $\frac{(P_x/P_z)^{^2\!\textrm{H}}}{(P_x/P_z)^{^1\!\textrm{H}}}$, calculated event-by-event, is also presented in
Fig.~\ref{fig:deuteronComponents}, using a procedure described in detail in the supplementary material.  As
expected from Fig.~\ref{fig:corrections}, the moving effect on the double ratio is small.

\begin{figure}[th]
\includegraphics[width=1\columnwidth]{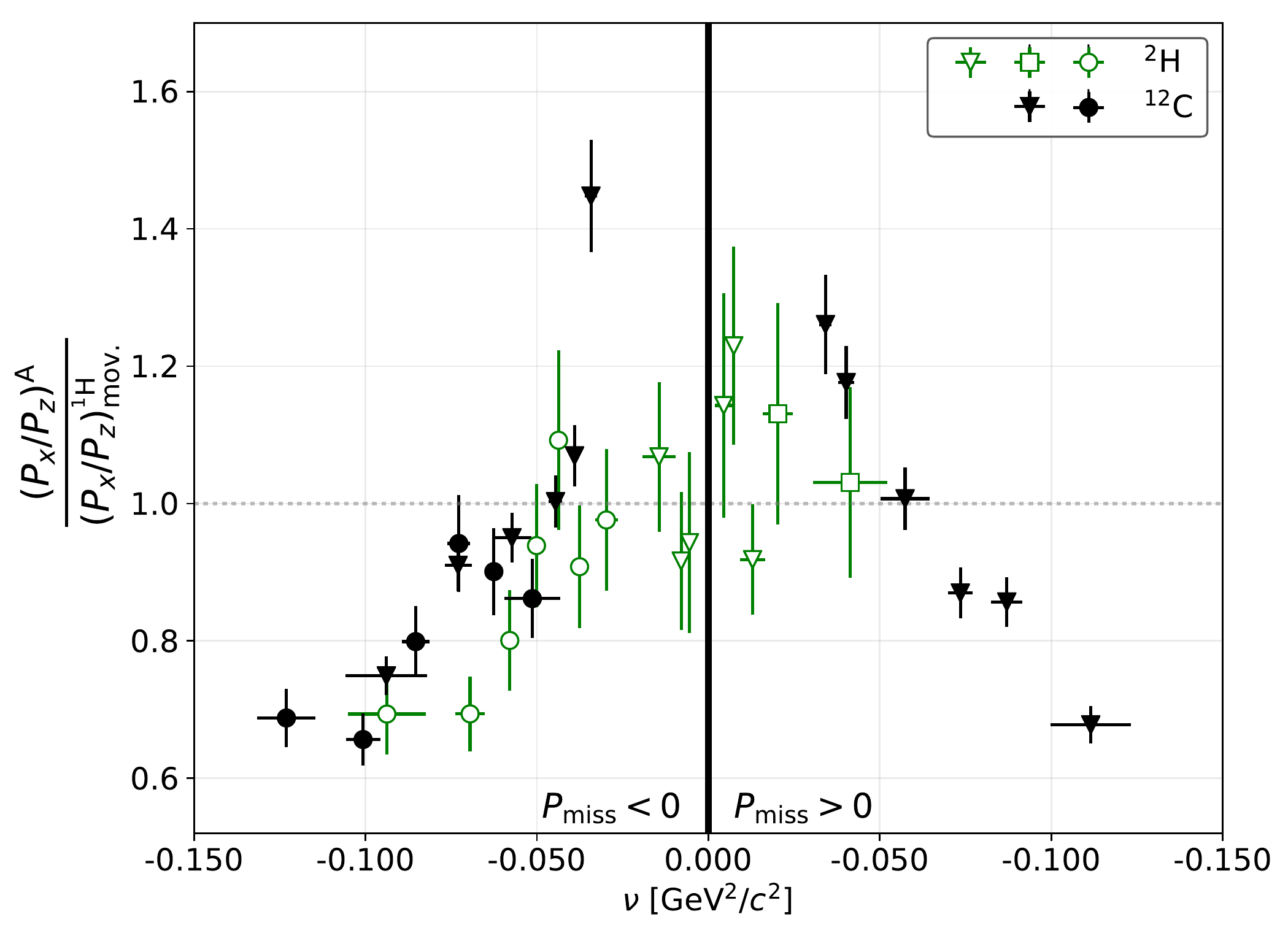}
\caption{The measured double ratios
$(P_x/P_z)^A/(P_x/P_z)^{^1\!\textrm{H}}$, using moving free-proton kinematics in the denominator, are shown as functions of the proton virtuality, for both $^{12}$C and $^2$H.  (The large double-ratios for carbon at $|\nu|<0.04$ GeV$^2/c^2$ have been attributed \cite{ceepLet} to knock-out of $p_{3/2}$-shell protons at small $|p_{\rm miss}|$ \cite{Ryckebusch99,GiustiCarlotta,Dutta}.)}
\label{fig:Deep_Ceep}
\end{figure}

A good agreement was found between the double ratios $\frac{(P_x/P_z)^A}{(P_x/P_z)^{^1\!\textrm{H}}}$ measured for $^{12}$C and $^2$H as functions of virtuality using resting-proton kinematics \cite{ceepLet}.  In order to test if this agreement is preserved using \textit{moving}-proton kinematics, we calculated the double ratios for the $^{12}$C datasets from MAMI \cite{ceepLet} in the same manner as for $^2$H.  The double ratios for the two nuclei are compared with one another in Figure \ref{fig:Deep_Ceep}, showing that the good agreement is maintained. 

\section{Conclusions}
%We observe that the free-proton kinematics for an initially-moving proton reduces, but does not eliminate the deviations between the measured components of the polarization transfer in quasi-free scattering and those in elastic scattering.   
We observe that using the moving-proton kinematics described in this paper reduces, but does not eliminate, the difference between measured components of the polarization transfer in quasi-free scattering and those in elastic scattering.   
The deviations from the free-proton in the transverse ($P_x$) component become smaller (around 7\%) when using the moving kinematics.  Similarly,  the deviations in the longitudinal ($P_z$) component are reduced, but remain significantly large, up to 50\% (Fig.~\ref{fig:deuteronComponents}).  
Since using the moving proton kinematics has a similar effect on both components, the effect on the double ratio largely cancels out.  Thus the double ratio is less sensitive to the kinematics than the separate components.
This implies that the choice of elastic kinematics is not the primary cause of the deviations in $P_x/P_z$ between quasi-free and elastic scattering.  

% It was previously noted that the virtuality is a useful parameter for comparing the double ratio $\frac{(P_x/P_z)^A}{(P_x/P_z)^{^1\!\textrm{H}}}$ of different nuclei at different $Q^2$.   
 %We have shown that the double ratios are also not sensitive to the initial momentum of the struck proton, allowing the virtuality to continue to be a preferred observable to study differences between bound and free protons.
 
 It was previously noted that the virtuality is a useful parameter for comparing the double ratio $\frac{(P_x/P_z)^A}{(P_x/P_z)^{^1\!\textrm{H}}}$ of different nuclei at different $Q^2$.  Here we show that the double ratio has only a weak sensitivity to the initial momentum of the proton, which continues to make the double ratio a preferred observable for studying the differences between bound and free protons.

\section*{Acknowledgements}
This work is
supported by the Israel Science Foundation (Grant 390/15) of the Israel
Academy of Arts and Sciences, 
by the PAZY Foundation (grant 294/18),
by the Deutsche Forschungsgemeinschaft (Collaborative Research
Center 1044), by the Slovenian Research Agency (research core funding
No.~P1\textendash 0102), by the U.S. National Science Foundation
(PHY-1505615), and by the Croatian Science Foundation Project No.~8570.

\section*{Appendix A: Supplemental material}
Supplementary material related to this article can be found online at \href{https://doi.org/10.1016/j.physletb.2019.04.004}{\small{https://doi.org/10.1016/j.physletb.2019.04.004}}.
\section*{References}
\bibliographystyle{elsarticle-num}
\addcontentsline{toc}{section}{\refname}\small{\bibliography{comp}}

\begin{thebibliography}{10}
\expandafter\ifx\csname url\endcsname\relax
  \def\url#1{\texttt{#1}}\fi
\expandafter\ifx\csname urlprefix\endcsname\relax\def\urlprefix{URL }\fi
\expandafter\ifx\csname href\endcsname\relax
  \def\href#1#2{#2} \def\path#1{#1}\fi

\bibitem{Jones:1999rz}
M.~K. Jones, et~al., { $G_{Ep} / G_{Mp}$ ratio by polarization transfer in
  polarized $\vec ep\!\rightarrow \!e \vec p$}, Phys. Rev. Lett. 84 (2000)
  1398--1402.
\newblock \href {http://arxiv.org/abs/nucl-ex/9910005}
  {\path{arXiv:nucl-ex/9910005}}, \href
  {http://dx.doi.org/10.1103/PhysRevLett.84.1398}
  {\path{doi:10.1103/PhysRevLett.84.1398}}.

\bibitem{Gayou:2001qd}
O.~Gayou, et~al., {Measurement of $G_{Ep} / G_{Mp}$ in $\vec ep\rightarrow e
  \vec p$ to $Q^2$ = 5.6 GeV$^2$}, Phys. Rev. Lett. 88 (2002) 092301.
\newblock \href {http://arxiv.org/abs/nucl-ex/0111010}
  {\path{arXiv:nucl-ex/0111010}}, \href
  {http://dx.doi.org/10.1103/PhysRevLett.88.092301}
  {\path{doi:10.1103/PhysRevLett.88.092301}}.

\bibitem{Punjabi:2005wq}
V.~Punjabi, et~al., {Proton elastic form-factor ratios to Q$^2$ = 3.5-GeV$^2$
  by polarization transfer}, Phys. Rev. C 71 (2005) 055202, [Erratum: Phys.
  Rev.C71,069902(2005)].
\newblock \href {http://arxiv.org/abs/nucl-ex/0501018}
  {\path{arXiv:nucl-ex/0501018}}, \href
  {http://dx.doi.org/10.1103/PhysRevC.71.055202, 10.1103/PhysRevC.71.069902}
  {\path{doi:10.1103/PhysRevC.71.055202, 10.1103/PhysRevC.71.069902}}.

\bibitem{Milbrath:1997de}
B.~D. Milbrath, et~al., {A Comparison of polarization observables in electron
  scattering from the proton and deuteron}, Phys. Rev. Lett. 80 (1998)
  452--455, [Erratum: Phys. Rev. Lett. 82, 2221 (1999)].
\newblock \href {http://arxiv.org/abs/nucl-ex/9712006}
  {\path{arXiv:nucl-ex/9712006}}, \href
  {http://dx.doi.org/10.1103/PhysRevLett.80.452, 10.1103/PhysRevLett.82.2221}
  {\path{doi:10.1103/PhysRevLett.80.452, 10.1103/PhysRevLett.82.2221}}.

\bibitem{Barkhuff:1999xc}
D.~H. Barkhuff, et~al., {Measurement of recoil proton polarizations in the
  electrodisintegration of deuterium by polarized electrons}, Phys. Lett. B 470
  (1999) 39--44.
\newblock \href {http://dx.doi.org/10.1016/S0370-2693(99)01294-0}
  {\path{doi:10.1016/S0370-2693(99)01294-0}}.

\bibitem{Pospischil:2001pp}
T.~Pospischil, et~al., {Measurement of $G_{Ep}/G_{Mp}$ via polarization
  transfer at $Q^2$ = 0.4 (GeV$/c)^2$}, Eur. Phys. J. A 12 (2001) 125--127.
\newblock \href {http://dx.doi.org/10.1007/s100500170046}
  {\path{doi:10.1007/s100500170046}}.

\bibitem{PhysRevC.64.038202}
O.~Gayou, K.~Wijesooriya, et~al., Measurements of the elastic electromagnetic
  form factor ratio ${\ensuremath{\mu}}_{p}{G}_{\mathrm{ep}}{/G}_{\mathrm{mp}}$
  via polarization transfer, Phys. Rev. C 64 (2001) 038202.
\newblock \href {http://dx.doi.org/10.1103/PhysRevC.64.038202}
  {\path{doi:10.1103/PhysRevC.64.038202}}.

\bibitem{MACLACHLAN2006261}
G.~MacLachlan, et~al., The ratio of proton electromagnetic form factors via
  recoil polarimetry at $q^2$ =1.13 $({\rm gev}/c)^2$, Nucl. Phys. A 764
  (2006) 261 -- 273.
\newblock \href {http://dx.doi.org/10.1016/j.nuclphysa.2005.09.012}
  {\path{doi:10.1016/j.nuclphysa.2005.09.012}}.

\bibitem{PhysRevC.74.035201}
M.~K. Jones, et~al., Proton ${G}_{E}/{G}_{M}$ from beam-target asymmetry, Phys.
  Rev. C 74 (2006) 035201.
\newblock \href {http://dx.doi.org/10.1103/PhysRevC.74.035201}
  {\path{doi:10.1103/PhysRevC.74.035201}}.

\bibitem{Akh74}
A.~I. Akhiezer, M.~Rekalo,
  \href{http://refhub.elsevier.com/S0370-2693(17)30052-7/bib416B683734s1}{Polarization
  effects in the scattering of leptons by hadrons}, Sov. J. Part. Nucl. 4
  (1974) 277, [Fiz. Elem. Chast. Atom. Yadra 4, (1973) 662].

\bibitem{Perdrisat}
C.~F. Perdrisat, et~al., {Nucleon electromagnetic form factors}, Prog. Part.
  Nucl. Phys. 59 (2007) 694--764.
\newblock \href {http://dx.doi.org/10.1016/j.ppnp.2007.05.001}
  {\path{doi:10.1016/j.ppnp.2007.05.001}}.

\bibitem{Sargsian}
M.~M. Sargsian, et~al., {Hadrons in the nuclear medium}, J.\ Phys. G 29 (2003)
  R1.
\newblock \href {http://dx.doi.org/10.1088/0954-3899/29/3/201}
  {\path{doi:10.1088/0954-3899/29/3/201}}.

\bibitem{deep2012PLB}
I.~Yaron, D.~Izraeli, et~al., Polarization-transfer measurement to a
  large-virtuality bound proton in the deuteron, Phys.\ Lett.\ B 769 (2017)
  21--24.
\newblock \href {http://dx.doi.org/10.1016/j.physletb.2017.01.034}
  {\path{doi:10.1016/j.physletb.2017.01.034}}.

\bibitem{deepCompPLB}
D.~Izraeli, I.~Yaron, et~al., {Components of polarization-transfer to a bound
  proton in a deuteron measured by quasi-elastic electron scattering}, Phys.
  Lett. B 781 (2018) 107--111.
\newblock \href {http://arxiv.org/abs/1801.01306} {\path{arXiv:1801.01306}},
  \href {http://dx.doi.org/10.1016/j.physletb.2018.03.063}
  {\path{doi:10.1016/j.physletb.2018.03.063}}.

\bibitem{Arenhovel}
H.~Arenh{\"o}vel, W.~Leidemann, E.~L. Tomusiak, {General survey of polarization
  observables in deuteron electrodisintegration}, Eur.\ Phys.\ J. A 23 (2005)
  147--190.
\newblock \href {http://dx.doi.org/10.1140/epja/i2004-10061-5}
  {\path{doi:10.1140/epja/i2004-10061-5}}.

\bibitem{ceepLet}
D.~Izraeli, T.~Brecelj, et~al., {Measurement of polarization-transfer to bound
  protons in carbon and its virtuality dependence}, Phys. Lett. B 781 (2018)
  95--98.
\newblock \href {http://arxiv.org/abs/1711.09680} {\path{arXiv:1711.09680}},
  \href {http://dx.doi.org/10.1016/j.physletb.2018.03.027}
  {\path{doi:10.1016/j.physletb.2018.03.027}}.

\bibitem{jlabDeep}
B.~Hu, et~al., {Polarization transfer in the $^2$H($\vec e, e' \vec p$)n
  reaction up to $Q^2$ = 1.61 (GeV/c)$^2$}, Phys.\ Rev. C 73 (2006) 064004.
\newblock \href {http://dx.doi.org/10.1103/PhysRevC.73.064004}
  {\path{doi:10.1103/PhysRevC.73.064004}}.

\bibitem{Strauch}
S.~Strauch, et~al., {Polarization transfer in the ${^4}$\textsc{H}e$(\vec
  e,e'\vec p)$${^3}$\textsc{H} reaction up to Q${^2}$ = 2.6~(GeV/$c$)${^2}$},
  Phys. Rev. Lett. 91 (2003) 052301.
\newblock \href {http://dx.doi.org/10.1103/PhysRevLett.91.052301}
  {\path{doi:10.1103/PhysRevLett.91.052301}}.

\bibitem{Paolone}
M.~Paolone, S.~P. Malace, S.~Strauch, et~al., Polarization transfer in the
  $^{4}\mathrm{He}(\vec e,e'\vec p)^{3}\mathbf{H}$ reaction at ${Q}^{2}=0.8$
  and $1.3\text{ }\text{ }(\mathrm{GeV}/c{)}^{2}$, Phys. Rev. Lett. 105 (2010)
  072001.
\newblock \href {http://dx.doi.org/10.1103/PhysRevLett.105.072001}
  {\path{doi:10.1103/PhysRevLett.105.072001}}.

\bibitem{Dieterich}
S.~Dieterich, et~al., Polarization transfer in the ${^4}$\textsc{H}e$(\vec
  e,e'\vec p)$${^3}$\textsc{H} reaction, Phys.\ Lett.\ B 500~(1--2) (2001) 47
  -- 52.
\newblock \href {http://dx.doi.org/10.1016/S0370-2693(01)00052-1}
  {\path{doi:10.1016/S0370-2693(01)00052-1}}.

\bibitem{Malov_O16}
S.~Malov, et~al., Polarization transfer in the ${}^{16}\mathrm{O}(\vec e,e'\vec
  p)^{15}\mathrm{N}$ reaction, Phys.\ Rev.\ C 62 (2000) 057302.
\newblock \href {http://dx.doi.org/10.1103/PhysRevC.62.057302}
  {\path{doi:10.1103/PhysRevC.62.057302}}.

\bibitem{Bernauer}
J.~C. Bernauer, et~al., {Electric and magnetic form factors of the proton},
  Phys.\ Rev. C 90~(1) (2014) 015206.
\newblock \href {http://dx.doi.org/10.1103/PhysRevC.90.015206}
  {\path{doi:10.1103/PhysRevC.90.015206}}.

\bibitem{ArenhovelMoving}
H.~Arenh{\"o}vel, {Polarization observables for elastic electron scattering off
  a moving nucleon}.~Submitted to Phys. Rev. C.
\newblock \href {http://arxiv.org/abs/1904.04515} {\path{arXiv:1904.04515}}.

\bibitem{a1aparatus}
K.~Blomqvist, et~al., The three-spectrometer facility at {MAMI}, Nucl.\
  Instrum.\ and Meth.\ A 403~(2--3) (1998) 263 -- 301.
\newblock \href {http://dx.doi.org/10.1016/S0168-9002(97)01133-9}
  {\path{doi:10.1016/S0168-9002(97)01133-9}}.

\bibitem{Pospischil:2000pu}
T.~Pospischil, et~al., The focal plane proton-polarimeter for the
  3-spectrometer setup at {MAMI}, Nucl.\ Instrum.\ Methods.\ Phys.\ Res.,
  Sect.\ A 483~(3) (2002) 713 -- 725.
\newblock \href {http://dx.doi.org/10.1016/S0168-9002(01)01955-6}
  {\path{doi:10.1016/S0168-9002(01)01955-6}}.

\bibitem{Ryckebusch99}
J.~Ryckebusch, D.~Debruyne, W.~Van~Nespen, S.~Janssen, Meson and isobar degrees
  of freedom in $(\vec e,e'\vec p)$ reactions at
  $0.2<{Q}^{2}<0.8~{(\mathrm{GeV}/c)}^{2}$, Phys.\ Rev.\ C 60 (1999) 034604.
\newblock \href {http://dx.doi.org/10.1103/PhysRevC.60.034604}
  {\path{doi:10.1103/PhysRevC.60.034604}}.

\bibitem{GiustiCarlotta}
C.~Giusti, J.~Ryckebusch, private communication (2017).

\bibitem{Dutta}
D.~Dutta, et~al., Quasielastic $(e,e'p)$ reaction on $^{12}\mathrm{C}$,
  $^{56}\mathrm{Fe}$, and $^{197}\mathrm{Au}$, Phys.\ Rev.\ C 68 (2003) 064603.
\newblock \href {http://dx.doi.org/10.1103/PhysRevC.68.064603}
  {\path{doi:10.1103/PhysRevC.68.064603}}.

\end{thebibliography}


\begin{thebibliography}{1}
\expandafter\ifx\csname url\endcsname\relax
  \def\url#1{\texttt{#1}}\fi
\expandafter\ifx\csname urlprefix\endcsname\relax\def\urlprefix{URL }\fi
\expandafter\ifx\csname href\endcsname\relax
  \def\href#1#2{#2} \def\path#1{#1}\fi

\bibitem{ceepLet}
D.~Izraeli, T.~Brecelj, et~al., {Measurement of polarization-transfer to bound
  protons in carbon and its virtuality dependence}, Phys. Lett. B 781 (2018)
  95--98.
\newblock \href {http://arxiv.org/abs/1711.09680} {\path{arXiv:1711.09680}},
  \href {http://dx.doi.org/10.1016/j.physletb.2018.03.027}
  {\path{doi:10.1016/j.physletb.2018.03.027}}.

\bibitem{deep2012PLB}
I.~Yaron, D.~Izraeli, et~al., Polarization-transfer measurement to a
  large-virtuality bound proton in the deuteron, Phys.\ Lett.\ B 769 (2017)
  21--24.
\newblock \href {http://dx.doi.org/10.1016/j.physletb.2017.01.034}
  {\path{doi:10.1016/j.physletb.2017.01.034}}.

\bibitem{deepCompPLB}
D.~Izraeli, I.~Yaron, et~al., {Components of polarization-transfer to a bound
  proton in a deuteron measured by quasi-elastic electron scattering}, Phys.
  Lett. B 781 (2018) 107--111.
\newblock \href {http://arxiv.org/abs/1801.01306} {\path{arXiv:1801.01306}},
  \href {http://dx.doi.org/10.1016/j.physletb.2018.03.063}
  {\path{doi:10.1016/j.physletb.2018.03.063}}.

\bibitem{polar2}
D.~Izraeli, I.~Mardor, E.~O. Cohen, M.~Duer, T.~Y. Izraeli, I.~Korover,
  J.~Lichtenstadt, E.~Piasetzky, {Polar polarization: a new method for
  polarimetry analysis}, JINST 13~(07) (2018) P07029.
\newblock \href {http://arxiv.org/abs/1803.06729} {\path{arXiv:1803.06729}},
  \href {http://dx.doi.org/10.1088/1748-0221/13/07/P07029}
  {\path{doi:10.1088/1748-0221/13/07/P07029}}.

\end{thebibliography}

\clearpage{}

\newpage{}\newpage{}\newpage{}

\end{document}

\begin{frontmatter}{}

\title{The influence of Fermi motion on the comparison of the polarization
transfer to a proton \\in elastic $\vec ep$ and quasi-elastic $\vec eA$ scattering.\\
Supplemental Material.}

\author[TAU]{S.~Paul\corref{cor2}}

%\ead{davidizraeli@post.tau.ac.il}
\ead{sebouh.paul@gmail.com}

\author[JSI]{T.~Brecelj}

\author[Mainz]{H.~Arenh\"ovel}

\author[Mainz]{P.~Achenbach}

\author[TAU]{A.~Ashkenazi}

\author[JSI]{J.~Beri\v{c}i\v{c}}

\author[Mainz]{R.~B\"ohm}

\author[zagreb]{D.~Bosnar}

\author[TAU]{E.O.~Cohen}

\author[JSI]{L.~Debenjak }

\author[Mainz]{M.O.~Distler}

\author[Mainz]{A.~Esser}

\author[zagreb]{I.~Fri\v{s}\v{c}i\'{c}\fnref{mit}}

\author[Rutgers]{R.~Gilman}

\author[TAU]{D.~Izraeli}

\author[JSI]{T.~Kolar}

\author[TAU,nrc]{I.~Korover}

\author[TAU]{J.~Lichtenstadt}

\author[TAU,soreq]{I.~Mardor}

\author[Mainz]{H.~Merkel}

\author[Mainz]{D.G.~Middleton}

\author[JSI,Mainz,UL]{M.~Mihovilovi\v{c} }

\author[Mainz]{U.~M\"uller}

\author[TAU]{M.~Olivenboim}

\author[TAU]{E.~Piasetzky}

\author[Mainz]{J.~Pochodzalla}

\author[huji]{G.~Ron}

\author[Mainz]{B.S.~Schlimme}

\author[Mainz]{M.~Schoth}

\author[Mainz]{F.~Schulz}

\author[Mainz]{C.~Sfienti}

\author[UL,JSI]{S.~\v{S}irca}

\author[JSI]{S.~\v{S}tajner }

\author[USK]{S.~Strauch}

\author[Mainz]{M.~Thiel}

\author[Mainz]{A.~Tyukin}

\author[Mainz]{A.~Weber}

\author[TAU]{I.~Yaron}

\author{\\\textbf{(A1 Collaboration)}}

\cortext[cor2]{Corresponding author}

\fntext[mit]{Present address: MIT-LNS, Cambridge, MA 02139, USA.}

\address[TAU]{School of Physics and Astronomy, Tel Aviv University, Tel Aviv 69978,
Israel.}

\address[JSI]{Jo\v{z}ef Stefan Institute, 1000 Ljubljana, Slovenia.}

\address[Mainz]{Institut f\"ur Kernphysik, Johannes Gutenberg-Universit\"at, 55099
Mainz, Germany.}

\address[zagreb]{Department of Physics, Faculty of Science, University of Zagreb, HR-10000 Zagreb, Croatia.}

\address[Rutgers]{Rutgers, The State University of New Jersey, Piscataway, NJ 08855,
USA.}

\address[nrc]{Department of Physics, NRCN, P.O. Box 9001, Beer-Sheva 84190, Israel.}

\address[soreq]{Soreq NRC, Yavne 81800, Israel.}

\address[UL]{Faculty of Mathematics and Physics, University of Ljubljana, 1000
Ljubljana, Slovenia.}

\address[huji]{Racah Institute of Physics, Hebrew University of Jerusalem, Jerusalem
91904, Israel.}

\address[USK]{University of South Carolina, Columbia, South Carolina 29208, USA.}

\begin{abstract}

We provide in this supplement the derivation of the ``moving'' free-proton kinematics which we use in the paper to calculate the free-proton polarizations to compare our data with.  We also provide
the kinematics and details of the experimental set-ups, a comparison of the measured and elastic kinematics for the $^{12}$C data, and information about the extraction of the polarization ratios.

\end{abstract}
\date{\today}

\end{frontmatter}{}
%define what subscript to use for the moving proton kinematics
\newcommand{\mov}{M}
\section{Derivation of moving proton kinematics}
Our goal here is to find a set of elastic $ep\rightarrow e'p'$ kinematics with the same initial electron energy $k_0$, momentum transfer invariant $Q^2 = -(k-k')^2$, and initial proton momentum $\vec{p}$ as in the measured quasi-free scattering reaction.  Here we assume there are no final-state interations (FSI) and therefore $\vec p = \vec p^{\,\prime}-\vec q = -\vec p_{\rm miss}$.  By definition, in the elastic kinematics, both the proton and the electron are on-shell in their initial and final states.  

The initial and final electron four-momenta are denoted $k$ and $k'$, while the initial and final proton momenta are denoted $p$ and $p'$.  The momentum transfer is denoted by $q = k-k'$ which, due to conservation of momentum, is also equal to $p'-p$.   Quantities without subscripts are from the measured kinematics.  A subscript $\mov$ denotes the quantities in the free moving-proton kinematics.  

In this derivation, the electron mass is neglected, and the coordinate system is chosen so that the momentum transfer $\vec{q}$ is in the $z$ direction and that $\vec{k}- (\vec{k} \cdot \hat{z}) \hat{z}$ is in the $x$ direction and the $y$ direction is perpendicular to both $x$ and $z$.  

First we note that the four momentum of the initial on-shell proton is trivially derived by taking the four-momentum from the measured event and adjusting its energy to put it on-shell:
\begin{equation}
p_\mov = \left(p_{0\mov}, \;p_x, \;p_y, \; p_z \right),
\end{equation}
where $p_{0\mov} = \sqrt{M^2+|\vec p|^2}$ and $M$ is proton mass.  The final proton momentum is determined by the conservation of four-momentum:
\begin{equation}
p'_\mov = \left(p_{0\mov}+ \omega_\mov, \; p_x, \;p_y, \;p_z + q_{z\mov} \right),
\label{eq:fin_p}
\end{equation}
where $q_{z\mov}$ and $\omega_\mov$, are the space-like and time-like components of the four-momentum transfer, $q_\mov$. We begin solving for these using the Bjorken condition:
\begin{equation}
1 = \frac{Q^2_\mov}{2p_\mov\cdot q_\mov},
\end{equation}
or equivalently, 
\begin{align}
0 &=  2p_\mov\cdot q_\mov - Q^2 \\
   &= 2p_{0\mov}\omega_\mov  - 2p_z q_{z\mov} - Q^2.
\end{align}
Substituting $q_{z\mov} = \sqrt{Q^2+\omega_\mov^2}$ yields
\begin{equation}
0 = 2\omega_\mov p_{0\mov}-2p_z\sqrt{Q^2+\omega_\mov^2}- Q^2.
\label{eq:nasty_equation_to_solve}
\end{equation}
Solving  Eq. \ref{eq:nasty_equation_to_solve} for $\omega_\mov$ yields:
\begin{equation}
  \omega_\mov =  \frac{Q^2 p_{0\mov}+p_z Q\sqrt{4E_{pt}^2+Q^2}}{2E_{pt}^2},
 \label{eq:omega_mov_deriv}
 \end{equation}
 where $E_{pt}$ is defined as $\sqrt{M^2+p_x^2+p_y^2}$.    In the coordinate system we have chosen, the lepton kinematics must be of the form:
 \begin{equation}
 k_\mov=(k_{0}, k_{x\mov}, 0, k_{z\mov}),
 \end{equation}
 and 
 \begin{equation}
 k'_\mov=(k_{0}-\omega_\mov, k_{x\mov}, 0, k_{z\mov}-q_{z\mov}),
 \end{equation}
 using the same $\omega_\mov$ and $q_{z\mov}$ previously solved for.  We have already chosen to make $k_{0\mov} = k_0$.  
 First, we enforce that the initial electron is on-shell\footnote{neglecting the electron mass}:
  \begin{equation}
 0 = k^2 = k_0^2-k_{x\mov}^2-k_{z\mov}^2,
 \label{kxm}
   \end{equation}
   which yields
    \begin{equation}
 k_{x\mov}=\sqrt{k_{0\mov}^2-k_{z\mov}^2}.
   \end{equation}
To obtain $k_{z\mov}$, we require that the final electron is also on-shell, yielding
 \begin{align}
 0 &= k'^2 \\
    &= (k_0-\omega_\mov)^2-k_{x\mov}^2-(k_{z\mov}-q_{z\mov})^2\\
     &= -2k_0\omega_\mov+2q_{z\mov} k_{z\mov} -Q^2,
 \end{align}
 thus 
 \begin{equation}
 k_{z\mov} = \frac{2k_0\omega_\mov+Q^2}{2q_{z\mov}}.
 \label{eq:kzm}
 \end{equation}
 Finally, substituting Equation \ref{eq:kzm} into Equation \ref{kxm} yields:
 \begin{align}
 k_{x\mov} =& \sqrt{k_0^2-\left(\frac{2k_0\omega_\mov+Q^2}{2q_{z\mov}}\right)^2},
 \end{align}
 which simplifies to
 \begin{align}
        k_{x\mov}=& \frac{Q\sqrt{4k_0(k_0-\omega_\mov)-Q^2}}{2q_{z\mov}},
 \end{align}
 where $\omega_\mov$ is given in Equation \ref{eq:omega_mov_deriv} and  $q_{z\mov} = \sqrt{Q^2+\omega_\mov^2}$.  Thus all of the four-vector components of the moving-proton kinematics in Section 2 of the letter have been derived.   Substituting $\vec{p} = 0$ yields the kinematics for a resting proton.

\section{Experimental setup details}
\begin{figure}[th!]
\begin{center}
\includegraphics[width=\columnwidth]{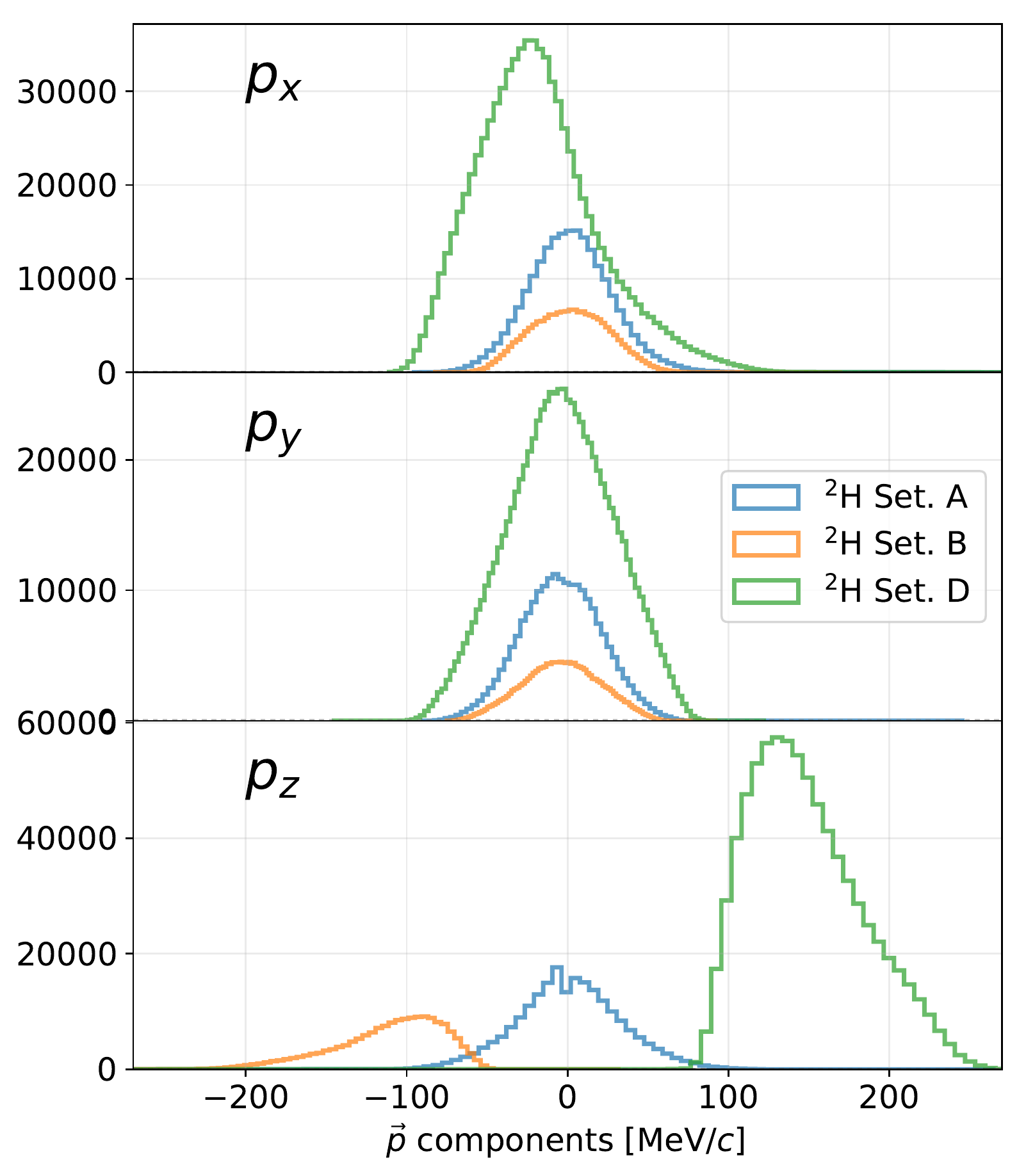}
\caption{
The spectra of the vector components of the proton initial momentum $\vec p$ of the measured events in the MAMI $^2$H datasets, assuming no FSI ($\vec{p}=-\vec{p}_{\rm miss}$).  Here, $\hat z$ is in the direction of the momentum transfer $\vec q$; $\hat x$ is in the scattering plane perpendicular to $\hat z$; and $\hat y$ is perpendicular to the scattering plane.
}
\label{fig:initial_momentum_d}
\end{center}
\end{figure}

The kinematic settings of the $^2$H($\vec{e},e'\vec p$)n and $^{12}$C($\vec{e},e'\vec p$) experiments at MAMI are listed in Tables \ref{tab:kinematics_d} and \ref{tab:kinematics_C}.  The spectra of the vector components of the proton initial momenta $\vec{p} = -\vec{p}_{\rm miss}$ for the $^2$H data are shown in Figure \ref{fig:initial_momentum_d}.  

\begin{table}[htp]
\caption{
The kinematic settings in the  $^2$H($\vec{e},e'\vec p$) experiment. The angles and momenta represent the central values for the two spectrometers: $p_p$ and $\theta_p$ ($p_e$ and $\theta_e$) are the knocked out proton (scattered electron) momentum and scattering angles, respectively.
Also shown are the ranges for $\theta_{pq}$, which is the angle between $\vec{q}$ and $\vec{p}$.}
\begin{center}
\begin{tabular}{llllll}
\hline
\multicolumn{2}{c}{Kinematic} & Setup & & & \\ \cline{3-6}
 & &  A & B  & D\\
\hline
$E_{\rm beam}$ & [MeV] & 600 & 600 & 630\\
$Q^2$ & [GeV$^2$] & 0.40 & 0.40  & 0.18\\
$p_{\rm miss}$ & [MeV] & -80 to 75 & 75 to 175  & -220 to -130 \\
$p_e$ & [MeV] & 384 & 463  & 398 \\
$\theta_e$ & [deg] & 82.4 & 73.8  & 49.4 \\
$p_p$ & [MeV] & 668 & 495  & 665 \\
$\theta_{p}$ & [MeV] & -34.7 & -43.3  & -39.1 \\
$\theta_{pq}$ & [deg] & 0 to 180 & 140 to 180  & 0 to 45 \\
\multicolumn{2}{l}{\tiny{\# of events after cuts}} & 213 k & 172 k  & 790 k \\
\hline
\end{tabular}
\end{center}
\label{tab:kinematics_d}
\end{table}%

\FloatBarrier
\begin{table}[ht!]
\caption{
The kinematic settings in the $^{12}$C($\vec{e},e'\vec p$) experiment. The angles and momenta represent the central values for the two spectrometers: $p_p$ and $\theta_p$ ($p_e$ and $\theta_e$) are the knocked out proton (scattered electron) momentum and scattering angles, respectively.  Also shown are the ranges for $\theta_{pq}$, which is the angle between $\vec{q}$ and $\vec p$.}
\begin{center}
\begin{tabular}{lllll}
\hline
\multicolumn{2}{c}{Kinematic} & Setup & & \\ \cline{3-5}
 & &  A & B \\
\hline
$E_{\rm beam}$ & [MeV] & 600 & 600 \\
$Q^2$ & [GeV$^2$] & 0.40  & 0.18\\
$p_{\rm miss}$ & [MeV] & $-$130 to 100  &$-$250 to $-$100 \\
$p_e$ & [MeV] & 385 & 368 \\
$\theta_e$ & [deg] & 82.4 & 52.9 \\
$p_p$ & [MeV] & 668 & 665 \\
$\theta_{p}$ & [MeV] & -34.7  & -37.8 \\
$\theta_{pq}$ & [deg] & 0 to 180  & 0 to 45 \\
\multicolumn{2}{l}{\tiny{\# of events after cuts}} & 1.7 M &  1.1 M \\
\hline
\end{tabular}
\end{center}
\label{tab:kinematics_C}
\end{table}

\FloatBarrier
\section{Kinematic variables in the $^{12}$C datasets}
\begin{figure}[t!]
\begin{center}
\includegraphics[width=\columnwidth, trim={0 0 0 0}, clip]{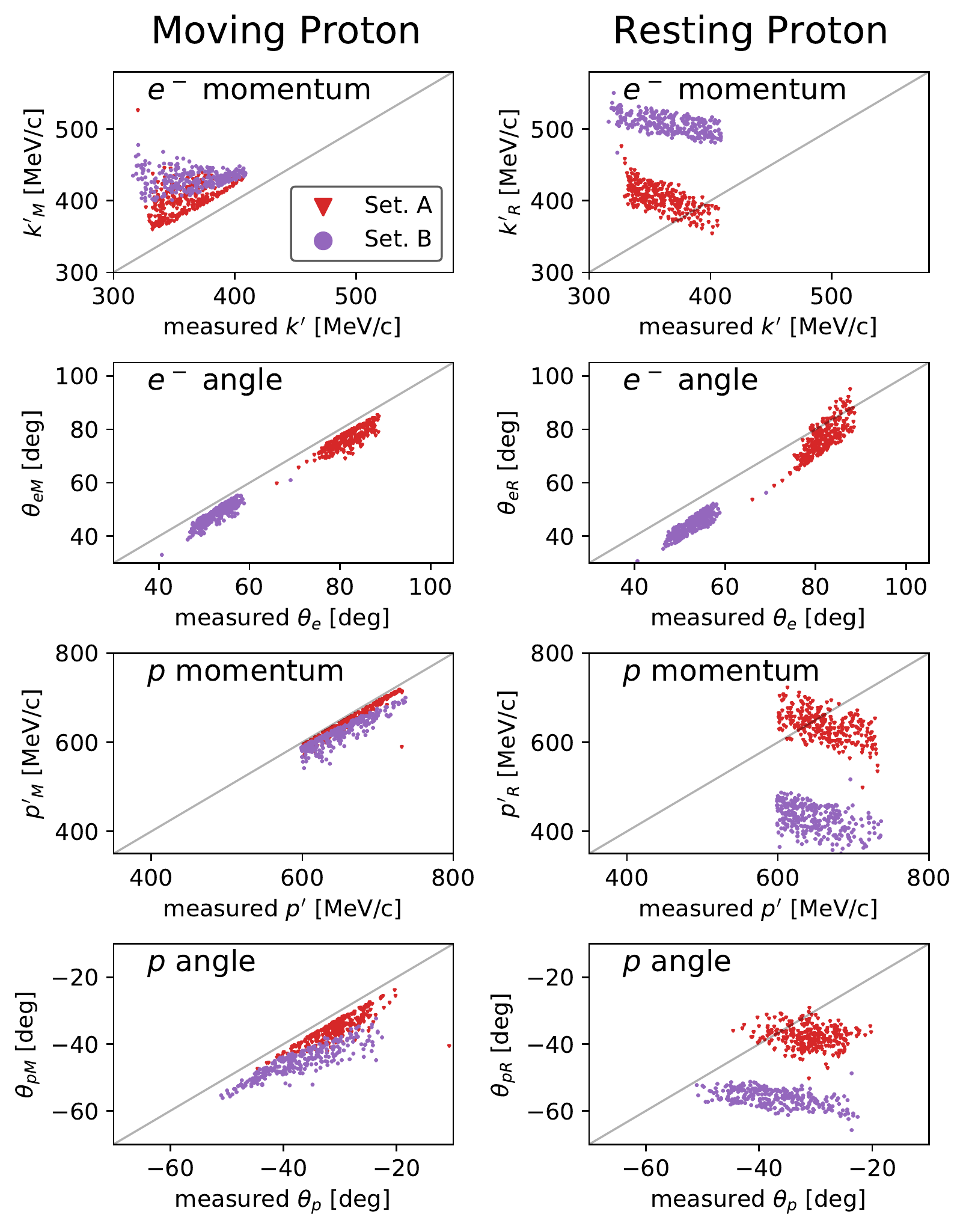}

\caption{
Comparison of the kinematic variables measured for $^{12}$C in \cite{ceepLet} ($x $-axis) to those in the moving-proton (left) and resting-proton (right) prescriptions ($y$-axis).  The different colors represent different kinematic settings.  The variables, from top to bottom are the scattered electron momentum ($k'$), scattered electron angle ($\theta_e$), recoiling proton momentum ($p'$) and recoiling proton angle ($\theta_p$).  The two angles are measured relative to the beam $\vec k$.  These variables were selected because they are the ones that are directly measured in the experiment.  
}
\label{fig:variables}
\end{center}
\end{figure}

In the $^{12}$C datasets \cite{ceepLet}, there are larger differences between the moving-proton kinematics and the measured kinematics than those in the $^2$H datasets \cite{deep2012PLB,deepCompPLB}, due to the larger Fermi momentum in the carbon compared to the deuteron.  However, this discrepancy is still much smaller than the discrepancy between the resting-proton and measured kinematics.  This can be seen in Figure \ref{fig:variables}, which, compares selected kinematic variables in the measured kinematics for the $^{12}$C datasets with those in the moving and resting proton kinematics, similar to Fig.~2 of the letter, which shows the $^2$H datasets.  

\section{Extraction of polarization-transfer ratios}
The extraction of the polarization ratios, as presented in the paper, are performed using the framework laid out in \cite{polar2}.  We find parameters that maximize the likelihood:
maximize the fit likelihood:
\begin{equation}
\mathcal{L} = \sum_i \left[\log(1+\vec{P}_i\cdot\vec{\lambda}_i)\right],
\end{equation}
where $\vec{\lambda}_i$ is the vector analyzing power of an event, determined by the secondary scattering angle, the precession and the beam polarization \cite{polar2}.  $\vec{P}_i$ are modeled polarizations for each event, characterized by some set of model-dependent parameters.  We employed two types of models for $\vec{P}_i$, and performed the fits separately for each bin in virtuality.  

The first type of model is the same one that we used in \cite{deepCompPLB}.  In this model we modeled the polarization for each event as:
\begin{equation}
P_i = \left(f_x P_{xi}^{^1\!\rm H},\; P_y,\; f_z P_{zi}^{^1\!\rm H}\right),
\end{equation} where $P_{xi}^{^1\!\rm H}$ and $P_{zi}^{^1\!\rm H}$ are the free-proton polarizations calculated event by event.  The free parameters of this model are $f_x$, $P_y$ and $f_z$.  The polarization ratios $\frac{(P_x)^A}{(P_x)^{^1\!\rm H}}$ and $\frac{(P_x)^A}{(P_x)^{^1\!\rm H}}$, shown in the top and middle panels of Figure 4 of the letter, are $f_x$ and $f_z$.  

The second type of model was designed to calculate the double ratio $\frac{(P_x/P_z)^A}{(P_x/P_z)^{^1\!\rm H}}$.  The modeled polarization for each event in the second model is
\begin{equation}
P_i = \left(r m \frac{P_{xi}^{^1\!\rm H}}{|P_i^{^1\!\rm H}|},\; P_y,\; m \frac{P_{zi}^{^1\!\rm H}}{|P_i^{^1\!\rm H}|}\right),
\end{equation} 
where $|P_i^{^1\!\rm H}| = \sqrt{(P_{xi}^{^1\!\rm H})^2 + (P_{zi}^{^1\!\rm H})^2}$.  The free parameters of this model are $r$, $P_y$ and $m$.   The advantages to using this model are that it obtains the double ratio directly as the parameter $r$, and that it is independent of the magnitude of the free-proton polarization.  In the letter, the fitted values for $r$ are shown in the bottom panel of Fig.~4 and also in Fig.~5.    
\section{References}
\bibliographystyle{elsarticle-num}
\addcontentsline{toc}{section}{\refname}\bibliography{comp}